# Erbium-doped nanoparticles in silica-based optical fibres

## Wilfried Blanc*, Valérie Mauroy, Bernard Dussardier

LPMC
Université de Nice-Sophia Antipolis, CNRS UMR6622, Parc Valrose, 06108 Nice cedex, France
Fax: +33 4 92 07 67 54     E-mail: wilfried.blanc@unice.fr, valerie.mauroy@unice.fr, bernard.dussardier@unice.fr
*Corresponding author

**Abstract:** Developing of new rare-earth (RE)-doped optical fibres for power amplifiers and lasers requires continuous improvements in the fibre spectroscopic properties (like shape and width of the gain curve, optical quantum efficiency, resistance to spectral hole burning and photodarkening,…). Silica glass as a host material for fibres has proved to be very attractive. However some potential applications of RE-doped fibres suffer from limitations in terms of spectroscopic properties resulting from clustering or inappropriate local environment when doped into silica. To this aim, we present a new route to modify some spectroscopic properties of rare-earth ions in silica-based fibers based on the incorporation of erbium ions in amorphous dielectric nanoparticles, grown *in-situ* in fiber preforms. By adding alkaline earth elements, in low concentration into silica, one can obtain a glass with an immiscibility gap. Then, phase separation occurs under an appropriate heat treatment. We investigated the role of three alkaline-earth elements: magnesium, calcium and strontium. We present the achieved stabilization of nanometric erbium-doped dielectric nanoparticles within the core of silica fibers. We present the nanoparticle dimensional characterization in fiber samples. We also show the spectroscopic characterisation of erbium in preform and fibre samples with different compositions. This new route could have important potentials in improving rare-earth doped fibre amplifiers and laser sources.

**Valerie Mauroy** obtained her Bachelor Degree in Physics from high school of Technology (Chalon-sur-Saone, France) in 2004 and her Masters Degree from School of Physics, The University of Orleans (Orleans, France) in 2009. She is currently working on her PhD Degree in nanostructuration of optical fibre by phase separation.

**Bernard Dussardier** is a Senior Scientist with CNRS (France), working at *Laboratoire de Physique de la Matière Condensée* (LPMC), University of Nice Sophia-Antipolis (UNS). He obtained his PhD from UNS on rare-earth doped optical fibres (1992). After a post-doc at the Optoelectronics Research Centre (Southampton, UK) working on new materials for amplifying optical fibres, has was appointed by CNRS at LPMC where he supervises the CNRS Optical Fibre Fabrication Facility and leads the Active Optical Fibre group. He has published 62 papers in international journals & proceedings, and 123 communications in conferences.

# 1  Introduction

Developing of new rare-earth (RE)-doped optical fibres for power amplifiers and lasers requires continuous improvements in the fibre's spectroscopic properties (like gain and efficiency characteristics, resistance to spectral hole burning and photodarkening,...) besides reduction in device size and economical efficiency. Silica glass as a host material for fibres has proved to be very attractive. However some potential applications of RE-doped fibres suffer from limitations in terms of spectroscopic properties resulting from clustering or inappropriate local environment when doped into silica. Several mixed-oxide and non-oxide alternatives to silica have been successfully proposed to obtain spectroscopically improved amplifying singlemode fibres, however inducing fabrication difficulties and high cost.

The route of interest here consists of using silica as a mechanical host and support of the fibre optical waveguide, and of embedding RE-ions within oxide nanoparticles of composition and structure different from those of silica, and small enough to induce acceptable scattering loss. When nanoparticles have a crystalline structure the term Transparent Glass Ceramics (TGC) is used [1], however they may also be amorphous, such as those obtained by phase separation [2]. In the following we use the term TGC in both cases for convenience.

Scarce reports on RE-doped TGC singlemode fibres use low melting mixed oxides prepared by a rod-in-tube technique [3], or mixed oxyfluorides using a double-crucible technique [4], both with a subsequent ceramming stage. However the low melting point of these materials causes low compatibility with silica components. Transition metal-doped silica-based TGC fibres were prepared by MCVD (Modified Chemical Vapour Deposition) and using a slurry method [5], i.e. the particles were synthesized before insertion into the silica tube-substrate. We have proposed a more straightforward technique allowing to embed RE ions within *in-situ* grown oxide nanoparticles in silica-based preforms [6]. The implemented principle is the spontaneous phase separation process [2] : Silicate systems can exhibit strong and stable immiscibility when they contain divalent metals oxides (MO, where M= Mg, Ca or Sr). For example, if a silicate glass containing few mol% MO is heated it will decompose into two phases : one silica-rich and one MO-rich in shape of spherical particles. Two key advantages of this process are that (i) nanoparticles are grown *in-situ* during the course of the fabrication process and (ii) there is no need (and associated risks) of nanoparticles manipulation by an operator. Further, the process takes advantage of the high compositional control and purity typical of the MCVD technique.

In the present paper, we aim at showing the potential of a simple technique to produce silica-based fibres with the core made of a spectroscopically-modified RE-doped TGC. We report on the growth of erbium-doped MO-silicate nanoparticles, that are stable after the fibre drawing stage using the MCVD and solution doping techniques. The nanoparticles mean diameter varies strongly with the composition. Losses as low as 0.4 dB/m at 1350 are reported, acceptable for fibre amplifier or laser applications. Further it is shown that the $Er^{3+}$ emission band is broadened when compared to a standard Er-doped fibre. This demonstrates the potentiality of this fabrication technique for applications such as fibre amplifiers.

## 2   Experimental details

Preforms were fabricated by the conventional MVCD technique. In this process, gaseous chlorides ($SiCl_4$, $GeCl_4$, $POCl_3$) are passed through a rotating silica tube, heated by an external burner in translation along the tube. Due to the high temperature, chlorides oxidize, forming particles which deposit on the inner wall. This porous layer turns into a glassy layer when the burner passes over it (the tube external temperature is around 1500 °C). In the final stage, the tube is collapsed into a rod at a temperature higher than 1800°C. Fibres were obtained by stretching preforms in a drawing tower at temperatures higher than 2000 °C under otherwise normal conditions. Opto-geometric properties of the perform (or fibre) are determined by adjusting the composition and number of layers. In our samples, phosphorous and germanium concentrations are ~1 mol% and 2 mol%, respectively.

Erbium and alkaline-earth ions were incorporated through the solution doping technique. An alcoholic solution (of desired strength of $ErCl_3.6H_2O$ or $MCl_2.6H_2O$) is soaked for two hours in the unsintered core layer. After removing the solution, the layer is dried and sintered. Erbium concentration is estimated through absorption spectra to be around 100 ppm. Three alkaline-earth concentrations in the solution were prepared: 0, 0.01 and 0.1 mol/l. The core diameter in the preforms was measured with a preform analyzer (York Technology P 101 ) to be 1 mm. It is about 8 μm in fibres.

## 3   Nanoparticles and attenuation characterizations

The fibres were characterized through scanning electron microscopy (SEM, backscattered electrons). One typical SEM picture from the exposed core section of a cleaved fibre is shown on Fig. 1. The nanoparticles are visible as bright spots, showing the important composition contrast compared to the silica background. The fibre core corresponds to the grey disk. The dark central part of the core is caused by the evaporation of germanium and alkaline-earth elements; this is a common artifact of the MCVD technique. It can be corrected by process optimization. The nanoparticles formation is linked to the addition of alkaline-earth ions.

Analysis of SEM images reveals that nanoparticle mean size depends on the alkaline-earth elements and its concentration. In the case of calcium and strontium incorporation, nanoparticle mean size is around 100 nm, and bigger nanoparticles are observed. This leads to a high level of scattering of light. In the case of magnesium incorporation, nanoparticles with a mean size of ~40 nm are obtained and the inter-particle distance is within the 100-500-nm range. No particle bigger than 100 nm was observed in these MgO-doped fibres, unlike the case of CaO and SrO-doped fibres. When the solution concentration increases from 0.1 to 1 mol/l, the density of particles remains nearly the same but the mean particle diameter almost doubles to ~ 80 nm. For laser and amplifier fibres, the most promising composition is based on the magnesium incorporation.

**Figure 1** SEM pictures of fibres doped with 0.1 mol/l of calcium.

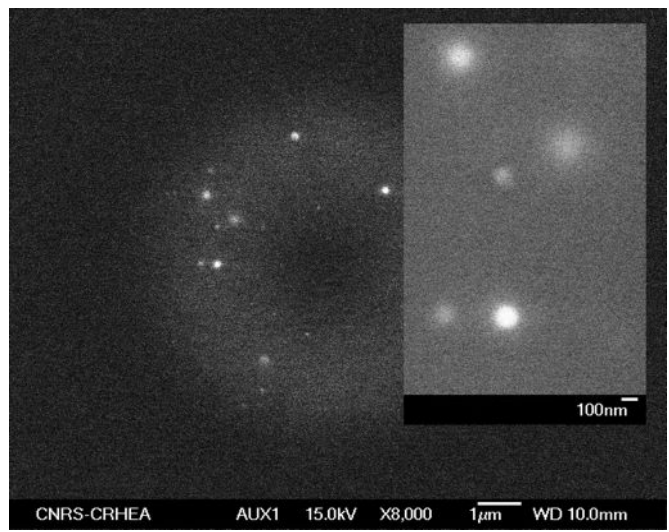

**Figure 2** SEM pictures of fibres doped with 0.1 mol/l of magnesium.

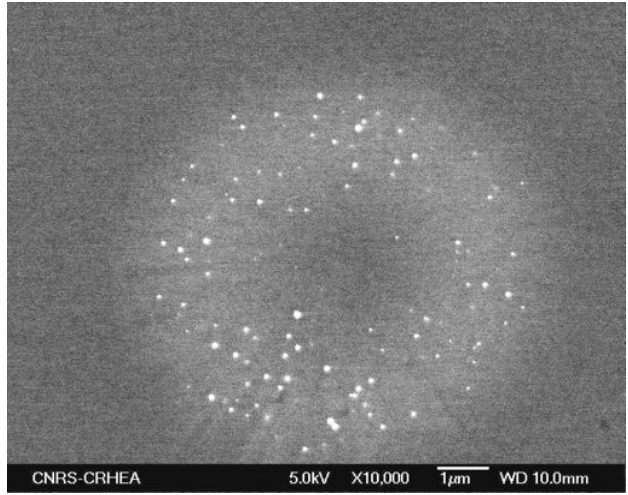

**Figure 3** SEM pictures of fibres doped with 0.1 mol/l of strontium

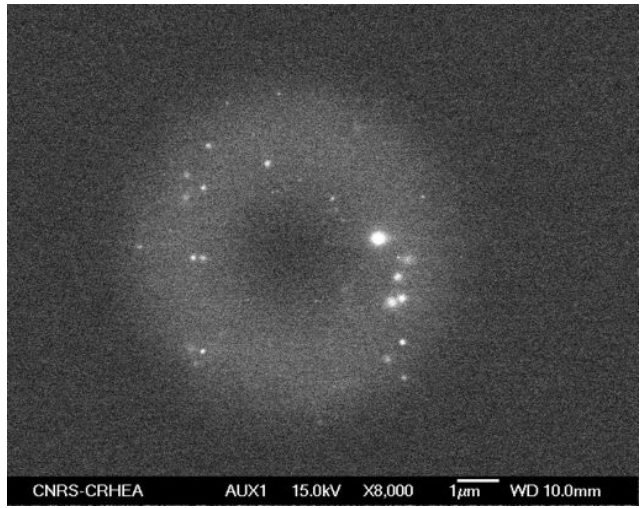

The scattering loss in Mg-doped fibre was measured through the cut-back method. The fibre bending radius was kept > 20 cm to minimize bend loss. The 0.1 mol/l Mg-doped fibre attenuation spectrum is displayed in Fig. 4. At the wavelength of 1350 nm, losses were measured to be 0.4 dB/m only. This value is comparable with the attenuation measured in low-melting temperature transparent glass ceramics fibres [7] and is compatible with amplifier applications. It is considered as a promising result.

**Figure 4** Transmission spectrum of a Mg-doped fiber (0.1 mol/l). Dots: experimental data, full line: calculated Rayleigh scattering curve.

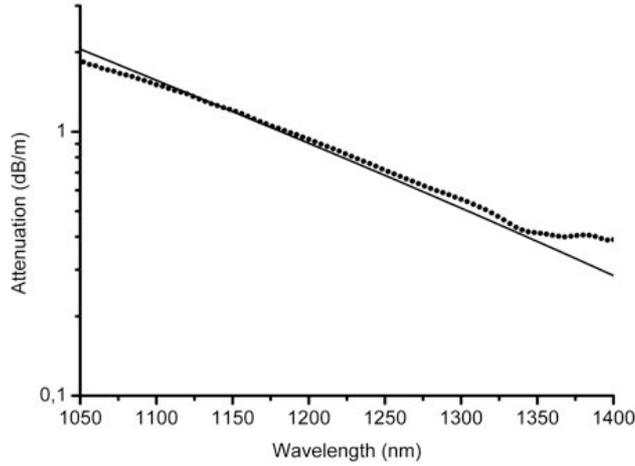

The Increase in loss with decrease in wavelength is attributable to stronger light scattering induced by the nanoparticles. Due to the small particle size we assume Rayleigh scattering to be the major source of scatter loss, which was estimated according to the formula: $\alpha(dB/m)=4.34.C_{Rayleigh}.N.\Gamma$, where $N$ is the DNP density ($m^{-3}$), $\Gamma$ is the overlap factor between the field and the core containing the DNP ($\Gamma = 0.3$ in fiber A) and $C_{Rayleigh}$ ($m^2$) is the Rayleigh scattering coefficient [8]:

$$C_{Rayleigh} = \frac{(2\pi)^5}{48} \times \frac{d^6}{\lambda^4} \times n_m^4 \times \left(\frac{n_n^2 - n_m^2}{n_n^2 + 2n_m^2}\right)^2$$

(1)

where $d$ is the nanoparticle diameter, $n_m$ and $n_n$ the host material and particles refractive indices, respectively. The actual particle composition is not known but a high content of MgO is expected [9]. The nanoparticle refractive index is estimated at ~ 1.65, like that of Mg-based oxide such as $Mg_2SiO_4$ [10]. Under these considerations, fitting of the experimental data with Eq. 1 yields a particle density $N\sim 0.4\ 10^{20}\ m^{-3}$, or equivalently a mean inter-particle distance ~300 nm. These values agree well with the results collected from the SEM pictures taken from cleaved fibres.

At 1350 nm, the normalized frequency $V$ was lower than 1.3 ($LP_{11}$ mode cut-off wavelength is 700 nm). The difference between experimental data and Rayleigh scattering curve above 1350 nm is attributed to bending loss only. However for practical application, the presented fabrication technique allows any necessary waveguide optimization without preventing the TGC growth. The attenuation of 1 mol/l Mg-doped fibre was extremely high (several 100s dB/m), in agreement with the Rayleigh formula (120 dB/m) assuming large monodisperse 100 nm particles. This shows that the particle mean size must be less than 50 nm for potential applications, and that the technique presented here is able to produce fibres with acceptable scattering loss.

## 4 Erbium spectroscopy

The composition of a Ca-doped sample was investigated by EDX analyses. When nanoparticles are analyzed, Ca, P and Si is found while only Si is detected outside of the particles. Germanium seems to be homogeneously distributed over the entire glass. When erbium is added to the composition, it is found to be inserted into the particles as it is presented on figure 5. Such conclusion was also drawn when emission intensity of $Er^{3+}$ was mapped on the end-face of a cleaved fibre under a confocal microscope [11]. No erbium fluorescence is detected outside the DNP. This result was expected due to the low solubility of silica (or even germano-silicate) for rare-earth ions, whereas strongly modified and amorphous silicates, such as in the DNP, have a high solubility for these ions. These results indicate that erbium ions are located inside or very close to the DNP.

**Figure 5** Energy Dispersive X-ray Analyses spectra of the preform sample doped with Ca and Er. (a) within a nanoparticle, (b) in the surrounding host, away from nanoparticles.

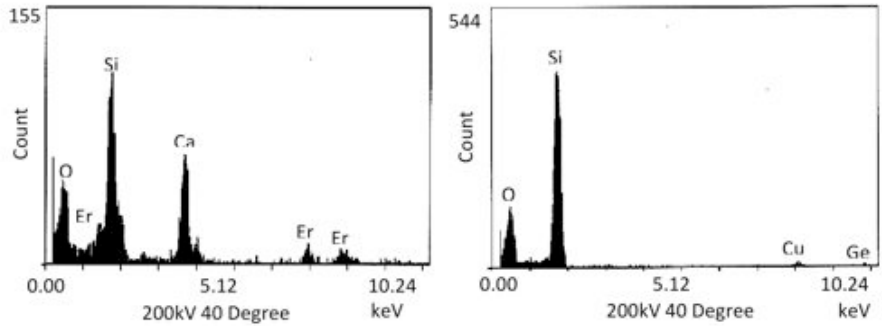

At room temperature, it was observed that erbium emission depends on calcium concentration. For the lowest concentration (0.01 mol/l), when no DNP are observed, erbium emission is similar to that from a standard silica erbium-doped preform where no calcium is added to the core composition (Figure 6). However, when calcium concentration is 0.1 mol/l, erbium emission spectrum is broadened (Figure 6). At very low temperature, the same was observed (not shown here). Such a broadening is also observed for different concentrations of Mg in fibres (figure 7).

**Figure 6** Room temperature erbium emission spectra in standard preform (dashed line) and DNP-doped preform with 0.01 and 0.1 mol/l of calcium (grey and black line, respectively). Excitation wavelength is 980 nm.

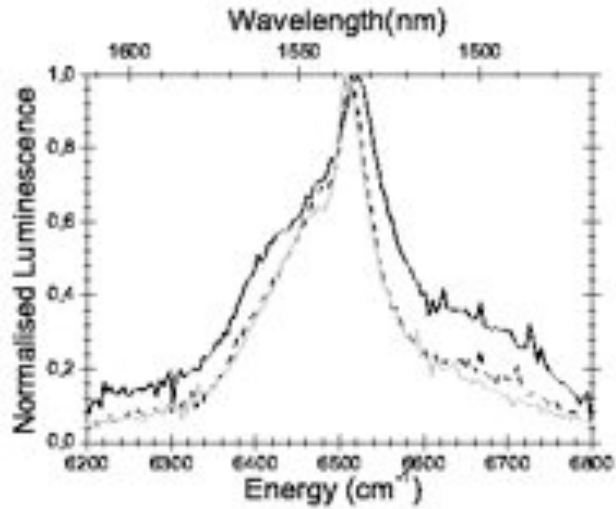

**Figure 7** Room temperature emission spectra from Mg-doped fibres. The Mg concentration in the doping solution is 0.1 (A) and 1.0 (B) mol/l, respectively. Excitation wavelength is 980 nm.

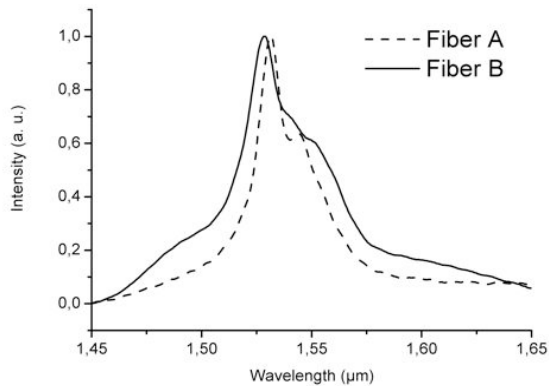

It is assumed that the change of local order around the erbium ions embedded into the alkaline-earth ions rich phase is responsible of the observed broadening. This luminescent broadening is the signature of a sensible modification of the close environment of rare earth ions in DNP-doped fibres, compared to standard silica fibres. Such assumptions were also supported by RFLN and EXAFS spectroscopy [11, 12].

## 5 Conclusion

A method to fabricate nanostructured $Er^{3+}$-doped fibres entirely through the MCVD process is demonstrated. By adding magnesium to the silica-based composition, nanoparticles of 40 nm in diameter are obtained through *in situ* growth without requiring a separate process to realize nanoparticles (such as post-process ceramming). Such fibres have low-loss. Moreover, a broadening of the emission spectrum by as much as almost 50% is observed with attractive features to realize gain flattened fibre amplifiers. More generally, this concept has great potentials as possible solutions to nowadays issues in amplifying fibres, including (but not limited to) amplifiers intrinsic gain flattening and/or spectral hole burning, amplifiers irradiation strengthening, photodarkening in fibre lasers, etc.

## Acknowledgements

This work was partly supported by CNRS (France), the Ministère des Affaires Etrangères (France) and Department of Science and Technology (India) through an Indo-French Research Network 'P2R' program. LPMC is a member of the GIS 'GRIFON' (http://www.unice.fr/GIS/). The authors thank Michèle Ude and Stanislaw Trzésien (LPMC, Nice, France) for samples preparation and Luan Nguyen (CRHEA, Valbonne, France) for SEM Analyses.